# Automatic Segmentation of Non-Tumor Tissues in Glioma MR Brain Images Using Deformable Registration with Partial Convolutional Networks

Zhongqiang Liu

Shanghai United Imaging Intelligence Co. Ltd, Shanghai, China , Southern Medical University

liuyear520@i.smu.edu.cn

**Abstract.** In brain tumor diagnosis and surgical planning, segmentation of tumor regions and accurate analysis of surrounding normal tissues are necessary for physicians. Pathological variability often renders difficulty to register a well-labeled normal atlas to such images and to automatic segment/label surrounding normal brain tissues. In this paper, we propose a new registration approach that first segments brain tumor using a U-Net and then simulates missed normal tissues within the tumor region using a partial convolutional network. Then, a standard normal brain atlas image is registered onto such tumor-removed images in order to segment/label the normal brain tissues. In this way, our new approach greatly reduces the effects of pathological variability in deformable registration and segments the normal tissues surrounding brain tumor well. In experiments, we used MICCAI BraTS2018 T1 tumor images to evaluate the proposed algorithm. By comparing direct registration with the proposed algorithm, the results showed that the Dice coefficient for gray matters was significantly improved for surrounding normal brain tissues.

## 1    Introduction

For brain tumor patients, precise segmentation of tumor region and quantitative assessment of surrounding normal tissues in MR images play important roles for diagnosis and surgical planning. Since tumors largely influence the brain morphology, it is difficult to accurately label the whole brain especially for the normal tissues surround the tumor area. Existing brain labeling algorithms include multi-atlas-based methods [1]-[3] and deep learning-based methods [4]. However, these algorithms can hardly tackle the tumor images, since most of them are investigated for normal brain (or brain without significant morphological variation) and have not considered tumor effects. Therefore, accurate and robust segmentation and labeling of normal tissues surrounding tumor is still a challenging task.

Several methods have been proposed to register pathological brain images with a normal brain atlas. In [5], a feature point-based radial mass-effect model is used to simulate tumor in the atlas and warp it onto the patient image. Masks of pathologies are utilized in [6] so that the warping close to a tumor is only driven by the information of the neighboring structures without tumor. However, the tumor structures vary considerably across patients in terms of location, shape, size, and extension, rendering difficulty in tumor simulation or masking.

In computer vision it has been shown that recovering the appearance inside a corrupted region is possible by using certain statistics and patterns from the remaining image via deep learning from a large number of samples. With convolutional neural network algorithms such inpainting amazingly restores the cut-out regions in terms of image appearance, contrast, and even shape [7][8]. Although only applied to regular corrupted regions such as rectangle, they can be extended to handle the inpainting of any irregular region. Motivated by this technique, it is believed that if the tumor area within a brain MR image is segmented and removed, such inpainting could recover the missing brain tissues, and the registration could be less affected by pathology, thereby giving relatively accurate brain labeling result especially to the surrounding normal tissues of that tumor. Since the location, size, and shape of each brain tumor is different, this kind of method can greatly reduce the tumor effects in MR brain image registration.

Therefore, in this paper, we propose to use a partial convolution network (PConv-Net) [9] to repair the brain tumor areas from the remaining normal tissues in brain MR images and then register the atlas image onto the subject image. The algorithm first uses a U-Net to segment the tumor, so that tumors with various sizes and locations can be removed first from the MR images. Then, a PConv-Net is applied to simulate normal tissue images with the cut tumor regions. Finally, we perform registration to register the atlas onto each recovered subject image to get more reasonable brain labeling result.

Our contributions include: a) a framework for brain tumor image registration; b) a segmentation model to segment brain tumors with various size and locations; c) a recovering network based on PConv-Net to recover normal tissue images from brain tumors before applying image registration.

To evaluate performance of the registration algorithm, we used the T1 and FLAIR images from MICCAI BraTS2018 in the experiments. Our method not only adapts to tumor areas with different shapes and sizes but also reduces the computational cost using partial convolutional operation. The registration results from our proposed method can be used to determine the tumor location and provide segmentation of the surrounding normal tissues to assist clinical diagnosis and treatment. Specifically, we used 100 cases for training and 40 cases for testing. The segmentation of BraTS2018 MR FLAIR was used as the ground-truth for segmentation on the corresponding T1 image for training the U-Net tumor segmentation. For training the PConv-Net, a tumor region from one subject was used as that of a different subject if the covered regions are normal. In this way, the ground truth for tumor region recovery using PConv-Net is obtained. After training the PConv-Net, we randomly selected 40 cases from our in-house normal brain MR images to test the repair performance of our PConv network. Finally, the 40 testing images were used for tumor segmentation, tumor region recovery, and performing registration to evaluate performance of the proposed algorithm.

## 2 Method

The framework of the proposed approach is illustrated in Fig. 1. In Fig. 1(a), a U-Net [10] is trained for tumor segmentation. In Fig. 1(b) a PConv-Net is trained to recover normal tissues within a given (tumor) region. Because the ground truth of the tumor region is unknown, to simulate the training data for PConv-Net, we applied artificial "tumor" regions on normal tissues of these images. After training, we can recover a normal image by two steps: 1) segment the tumor using the trained U-Net as shown in Fig. 1(a); 2) mask-out and recover the tumor region using PConv-Net as shown in Fig. 1(b). Finally, the SyN registration algorithm [11] was applied to register the atlas to the subject image for corresponding tissue segmentations and labels can be obtained.

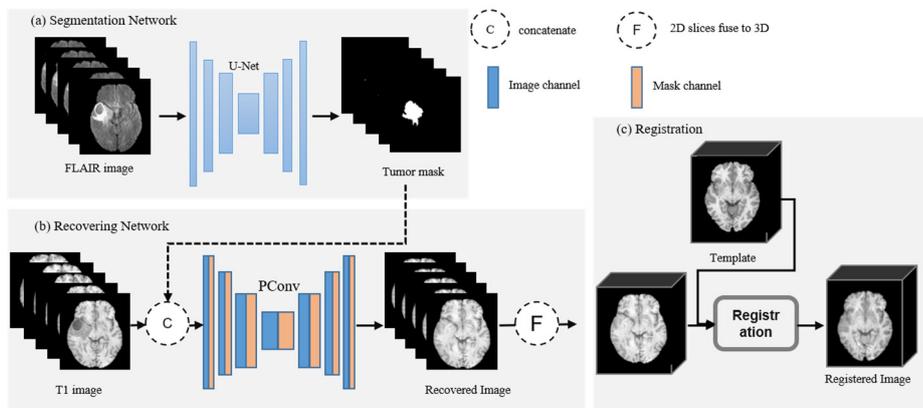

**Fig. 1.** The framework of the proposed method. (a) Tumor segmentation using U-Net; (b) partial convolutional network (PConv-Net) for tumor-region recovery; (c) registration.

### 2.1 Segmentation Network

The traditional U-Net is used for tumor segmentation. The U-Net structure, also known as encoder-decoder structure, is composed of two parts. The encoder path contains 3 down-sampling layers and 3 max-pooling layers used for feature extraction, and the decoder path includes 3 up-sampling layers for result recovery. In this step, in order to minimize tumor boundary intensity propagation effects, we expand the tumor mask using morphological operations before recovery (element unit size is 3mm). The segmentation was implemented based on 2D slices with smoothness constraints across neighboring slices, so the training time and memory are greatly reduced. We used the binary cross entropy loss (BCELoss) to measure reconstruction errors.

### 2.2 Recovery Network

**Partial Convolution.** To reduce the effect of tumor regions in image registration, a PConv-Net is applied to simulate normal tissue images to replace the tumor region. For this purpose, we adopted the partial convolution strategy from [12], where convolutional operations are only operated on a certain part of the image. Mathematically, one PConv layer can be expressed as:

$$x' = \begin{cases} F^T(X \odot M) \dfrac{sum(\mathbf{1})}{sum(M)} + b, \text{ if } sum(M) > 0 \\ 0 \qquad\qquad\qquad\quad, \text{ otherwise} \end{cases}, \qquad (1)$$

where $x'$ represents the output after convolution of the input image/feature map $X$. $F$ represents the weighting vectors of the convolution filter, and b is the bias vector. $M$ is a binary mask, which represents the tumor mask in our study." $\odot$ "denotes the element-wise multiplication, and 1 has the same shape as $M$ but with all elements being one. As can be seen, the output values depend only on the unmasked inputs as well as the shape of the mask. The scaling factor $sum(\mathbf{1})/sum(M)$ provides appropriate scaling to adjust the varying amount of valid (unmasked) inputs. In multi-layer convolution, the output from the previous layer $x'$ is used as the input $X$ of the next layer. No down-sampling and up-sampling are used in PConv-Net so that the mask $M$ has the same dimension with the input image and the feature maps.

**Mask Update.** After each partial convolution operation, we update the mask using Eq. (2). For the input mask $M$, the output $m'$ is actually a thresholding of the input:

$$m' = \begin{cases} 1, & \text{if } sum(M) > 0 \\ 0, & \text{otherwise} \end{cases}. \qquad (2)$$

This new mask is then fed to the next layer. As the number of network layers increases, the number of pixels in the mask output $m'$ decreases to 0, and area of the effective area in the output result $x'$ increases. The effect of mask on the overall loss will be smaller.

**Loss Functions.** The loss functions are aimed at both pixel-wise reconstruction accuracy and how smooth the repaired area transition into their surrounding tissue.

First, the MSE loss is used to ensure the recovered image patch is similar to the ground truth. Given an input image $I_{in}$ masked by the segmented tumor and the initial binary mask $M$ (0 for tumor area and 1 for other regions), the network prediction $I_{out}$ should be close to the ground-truth image $I_{gt}$, as defined by:

$$L_{masked} = \frac{1}{|I_{gt}|} \left\| (1 - M) \odot (I_{out} - I_{gt}) \right\|_1, \qquad (3)$$

$$L_{valid} = \frac{1}{|I_{gt}|} \left\| M \odot (I_{out} - I_{gt}) \right\|_1, \qquad (4)$$

where $|I_{gt}|$ denotes the number of elements in $I_{gt}$.

The perceptual loss $L_{perc}$ reflects the appearance of recovered areas and measures high-level perceptual and semantic differences between images, which is defined as:

$$L_{perc} = \sum_{p=0}^{P-1} \frac{\left\| \Psi_p^{I_{out}} - \Psi_p^{I_{gt}} \right\|_1}{\left| \Psi_p^{I_{gt}} \right|} + \sum_{p=0}^{P-1} \frac{\left\| \Psi_p^{I_{comp}} - \Psi_p^{I_{gt}} \right\|_1}{\left| \Psi_p^{I_{gt}} \right|}, \qquad (5)$$

where $I_{comp}$ is the raw output image $I_{out}$ but with the non-hole pixels directly set to the ground truth. $\left| \Psi_p^{I_{gt}} \right|$ is the number of elements in $\Psi_p^{I_{gt}}$, where $\Psi_p^{I}$ denotes the activation map of the $p$th layer of the network. The perceptual loss calculates the $L^1$ distance between $I_{out}$ and $I_{comp}$ based on autocorrelation (Gram matrix). The type loss term is introduced in each feature map:

$$L_{style\_out} = \sum_{p=0}^{P-1} \frac{1}{C_p C_p} \left\| K_p \left( \psi_p^{I_{out}} \right)^T \left( \psi_p^{I_{out}} \right) - \left( \psi_p^{I_{gt}} \right)^T \left( \psi_p^{I_{gt}} \right) \right\|_1, \quad (6)$$

$$L_{style\_comp} = \sum_{p=0}^{P-1} \frac{1}{C_p C_p} \left\| K_p \left( \psi_p^{I_{comp}} \right)^T \left( \psi_p^{I_{comp}} \right) - \left( \psi_p^{I_{gt}} \right)^T \left( \psi_p^{I_{gt}} \right) \right\|_1, \quad (7)$$

where $\Psi(x)_p$ represents the high level features of shape $(H_p W_p) \times C_p$, resulting in a $C_p \times C_p$ Gram matrix, $K_p$ is the normalization factor $\frac{1}{C_p H_p W_p}$ for the $p$th selected layer.

The final loss term is the total variation (TV) loss $L_{TV}$, which is the smoothing penalty on R, and R is the region of 1-pixel dilation of the recovered region:

$$L_{TV} = \sum_{(i,j) \in R, (i,j+1) \in R} \frac{\left\| I_{comp}^{i,j+1} - I_{comp}^{i,j} \right\|_1}{N_{I_{comp}}} + \sum_{(i,j) \in R, (i+1,j) \in R} \frac{\left\| I_{comp}^{i+1,j} - I_{comp}^{i,j} \right\|_1}{N_{I_{comp}}}, \quad (8)$$

where $N_{I_{comp}}$ is the number of elements in $I_{comp}$.

The total loss $L_{total}$ is the combination of all above loss functions [9]:

$$L_{total} = 6 L_{masked} + L_{valid} + 0.05 L_{perc} + 120 \left( L_{style\_out} + L_{style\_comp} \right) + 0.1 L_{TV} . \quad (9)$$

The architecture of PConv-Net is similar to U-Net. ReLU layer is used in encoder phase, and Leaky ReLU layer is used in decoder phase. The batch normalization (BN) layer is used in all layers except the first layer of encoder and the last layer of decoder.

## 3    Results

### 3.1    Data and Experimental Setting

We used the dataset from MICCAI BraTS2018, including 140 cases to evaluate the performance of the proposed algorithm. The T1-weighted and FLAIR images and their labels were used for tumor segmentation. Since MR scans often display inhomogeneity in intensity distribution caused by the bias fields, we first applied the N4 bias field correction [14] in the preprocessing. Then, to reduce the intensity distribution variability across different subjects, we first applied histogram matching and then normalized each image by subtracting the mean and dividing by the standard deviation. The T1 and FLAIR images are already registered, the tumor mask is extracted from FLAIR images.

To train the segmentation network, we resampled the FLAIR images and the label images into 240×240 with 155 slices. We randomly extracted 100 images for training, and then randomly select forty images in the remaining images for testing. The outputs of the network are corresponding tumor masks. For PConv-Net, 100 simulated images (arbitrarily putting simulated tumor masks on normal tissues) are used to train the model for tumor region recovery. Then we evaluated the performance of PConv-Net using 40 T1 images of normal subjects. The original T1 images and simulated tumor masks were used as the inputs and the tumor-recovered images are obtained. In this way the ground truth under the tumor masks is known. The PConv-Net was implemented using Pytorch with Adam optimization. The learning rate is initially set to 2e-4, then fine-tune using a learning rate of 5e-5. The maximum iterations are 1000k and batch size is 6. The network was trained in one GPU (12G NVIDIA GEFORCE RTX 2080 Ti). Finally, we stacked these repaired 2D slices into 3D images for registration.

### 3.2 Evaluation of Image Recovery

To evaluate the performance of image recovery, we generated tumor masks on normal images by removing the tissues within the masks and comparing the recovered images with the original images within those masks. Two metrics were used to compare the recovered image with the original image, including the peak signal-to-noise ratio (PSNR) and the structural similarity (SSIM) [13]. PSNR is measured in decibels (dB), and the higher PSNR generally indicates better reconstruction performance. SSIM ranges from 0 to 1, and 1 means perfect recovery.

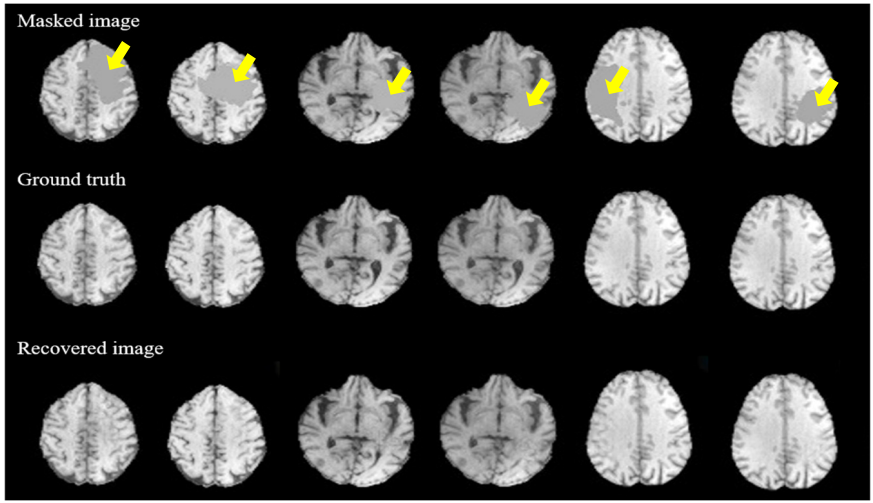

**Fig. 2.** Examples of recovered images within masks. Top: normal images with simulated masks (yellow arrows); middle: ground truth (original normal images); bottom: recovered images.

Fig. 2 shows some examples of image recovery result using PConv-Net. The first row shows the original images with generated masks, and the third row displays the recovered images. The original ground-truth images are shown in the second row. We can see that the mask-removed areas were recovered with appearance similar to the real ones. We performed quantitative evaluation by using T1 images with different masks, and obtained PSNR as 42.29±2.39dB and SSIM as 0.9989±6.3550E-05, indicating that our recovered image is very close to the original one.

Then, we applied the trained PConv-Net for the real tumor images. Fig. 3 shows two results. The first row is the original T1 images, and the segmented tumors are shown in the second row. By removing all the tumor regions, we can use PConv-Net to recover the images as shown in the third row. Because the tumor boundaries are often not obviously distinct with the surrounding tissues, we performed a dilation process (using an element unit with size 3mm) on the tumor mask so that the intensities of the tumor areas do not affect the recovery operation. Since the ground truth within the masked area is unknown, only qualitative results can be shown.

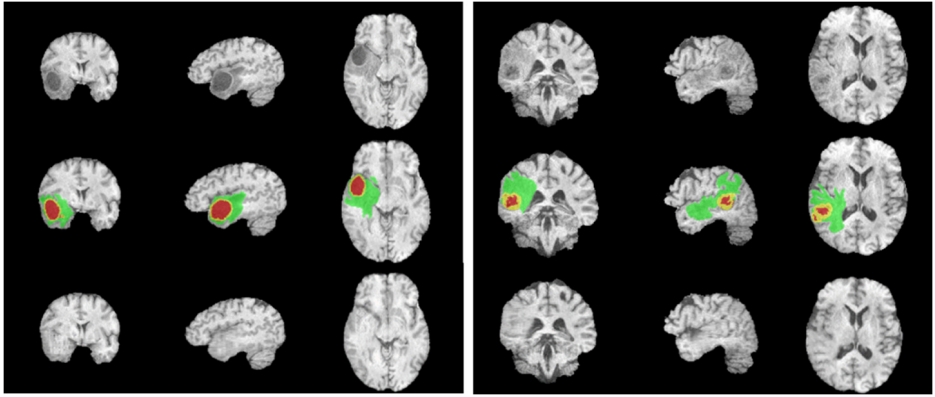

**Fig. 3.** Example results of PConv-Net for T1 images with glioma. Top: input image; middle: tumor segmented; bottom: tumor-recovered images.

### 3.3    Evaluation of Image Registration

Registration is performed to register the T1 template to each tumor-recovered subject image by using SyN. To evaluate the performance of image registration we used the resultant deformation fields to warp the segmentation of the atlas onto each subject and compared them with the manually corrected segmentation of each subject. Dice indices of grey matter (GM), whiter matter (WM) and cerebrospinal fluid (CSF) were computed for the evaluation. Dice coefficients are defined as $2|A_i \cap B_i|/|A_i| \cup |B_i|$, and $i$ represents the ROI label. A is deformed segmentation from the atlas image, B is the manually corrected segmentation on each subject.

For comparison, we computed the Dice coefficients outside the tumor regions, for evaluating the registration results. The registration is peformed with and without tumor region recovery using PConv-Net. Fig. 4 shows the comparison results. Fig. 4(a) is the input image, and Fig. 4(b) is the segmentation map warped from the atlas after registration. Fig. 4(c) is the recovery result of Fig. 4(a) using PConv-Net and Fig. 4(d) is the segmentation map obtained by registering the atlas to the recovered image.

In summary, for all the 40 testing subjects, we extract the tumors and compared Dice between the registered segmentation of the atlas and the manually corrected segmentation. An example of the input image and the segmentation images are shown in Fig. 5.

Table 1 shows the quantitative results over 40 testing images. It can be seen that Dice coefficients were significantly improved for GM (p<0.0005), while there is no significant differences for WM and CSF. Overall the proposed algorithm effectively achieved relatively accurate registration in terms of segmentation Dice.

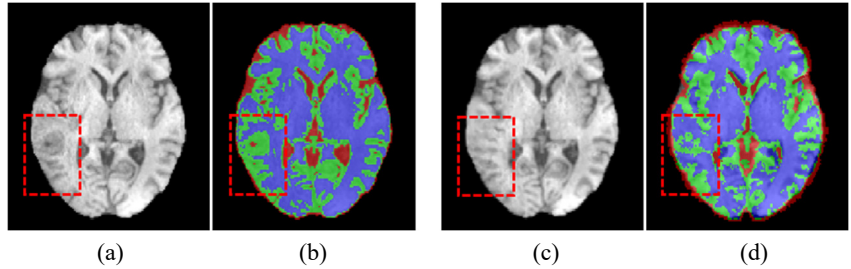

**Fig. 4.** A typical result of registration. (a) T1 image with tumor; (b) segmentation result (registering the atlas to the original images); (c) the recovered image using PConv-Net; (d) segmentation result (registering the atlas to the recovered images).

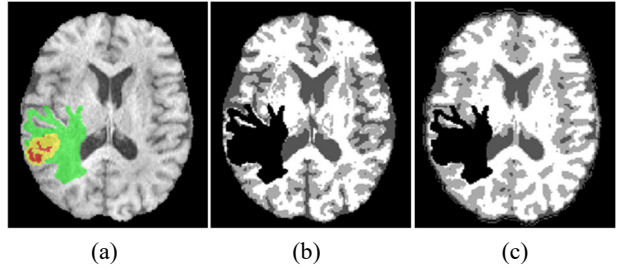

**Fig. 5.** An example of the segmentation. (a) Original image with tumor; (b) manually corrected segmentation; (c) warped segmentation map from the atlas.

**Table 1.** Dice of registration with and without PConv. "*" indicates significant improvement.

|  | CSF | GM* | WM |
| --- | --- | --- | --- |
| Dice within normal region without using PConv | 0.577 (±0.022) | 0.696(±0.041) | 0.793 (±0.027) |
| Dice within normal region using PConv | 0.569 (±0.015) | 0.724(±0.027) | 0.797 (±0.026) |

## 4    Conclusion

A brain MR image registration approach was proposed for registering a normal atlas onto the images with brain tumor. The algorithm is based on recovering missed normal tissues within the tumor regions using a partial convolutional network. Recent learning algorithms suggest that if the tumor area within an MR image can be recovered with normal tissues, the registration could be less affected by pathology, giving relatively accurate registration. Experiments with BraTS2018 tumor images showed the effectiveness of this strategy. We believe that with more accurate segmentation of the normal tissues of tumor image, better assessment or surgical planning can be performed. In the future work, we will further evaluate the performance of labeling the surrounding tissues and study the performance of glioma assessment with clinical datasets.